\documentclass[12pt]{article}                   
\usepackage{amssymb,amsmath}
\begin{document}
\vspace{10mm}
\begin{center}

\vspace{10mm}
\begin{center}

{\Large \bf The symmetry, inferable from Bogoliubov transformation, 
between the processes induced by the mirror in two-dimentional and the
charge in four-dimentional space-time}\\

 \vspace{10mm}  V.I.Ritus ${}^\dag$\\

\vspace{3mm}{$\dag$\it I.E.Tamm Department of Theoretical Physics,
              Lebedev Physical Institute,\\
 Leninsky Prospect 53, 119991, Moscow, Russia\\
             e-mail: ritus@lpi.ru}

    \vspace{2mm}

    \end{center}

    \end{center}
        \begin{abstract}

 The symmetry between the creation of pairs of massless bosons or
fermions by accelerated mirror in 1+1 space and the emission of single photons
or scalar quanta by electric or scalar charge in 3+1 space is deepened in this
paper. The relation of Bogoliubov coefficients, describing the processes
generated by the mirror, with Fourier's components of current or charge density
leads to the coicidence of the spin of any disturbances bilinear in scalar or
spinor field with the spin of quanta emitted by the electric or scalar charge.
The mass and invariant momentum transfer of these disturbances are essential for
the relation of Bogoliubov coefficients with invariant singular solutions and
Green's functions of wave equations both for 1+1 and 3+1 spaces and especially
for the integral relations (20) and (100) between these solutions. Namely the 
relation (20) leads to the coincidence of the self-action changes and 
vacuum-vacuum amplitudes for the accelerated mirror in two-dimentional 
space-time and charge in four-dimentional space-time. Thus, both invariants of 
the Lorentz group, spin and mass, perform intrinsic role in established symmetry.
The symmetry embraces not only the processes of real quanta radiation. It extends
also to the processes of the mirror and the charge interactions with the fields
carring spacelike momenta. These fields accompany their sources and define the
Bogoliubov matrix coefficients $\alpha^{B,F}_{\omega'\omega}$. It is shown that
the Lorentz-invariant traces $\pm {\rm tr}\,\alpha^{B,F}$ describe the vector 
and scalar interactions of accelerated mirror with uniformly moving detector.
This interpretation rests essentially on the relation (100) between the 
propagators of the waves with spacelike momenta in 2- and 4-dimentional spaces.
The traces $\pm {\rm tr}\,\alpha^{B,F}$ coincide actually with the products of 
the mass shift $\Delta m_{1,0}$ of accelerated electric or scalar charge and the 
proper time of the shift formation. The symmetry fixes the value of the bare
fine structure constant $\alpha_0=1/4\pi$.

\end{abstract}

\section{Introduction}
The Hawking's mechanism for particle production at the black hole formation is
analogous to the emission from an ideal mirror accelerated in vacuum [1].
In its turn there is a close analogy between the radiation of pairs of scalar
(spinor) quanta from accelerated mirror in 1+1 space and the radiation of
photons (scalar quanta) by an accelerated electric (scalar) charge in 3+1 space
[2,3]. Thus all these processes turn out to be mutually related. In problems
with moving mirrors the $in$- and $out$-sets of the wave equation solutions are
usually used which for massless scalar field look as follows
\begin{equation}
\phi_{in\,\omega'}\sim e^{-i\omega' v}-e^{-i\omega' f(u)},\quad \phi^*
_{in\,\omega'};\qquad
\phi_{out\,\omega}\sim e^{-i\omega g(v)}-e^{-i\omega u},\quad \phi^*
_{out\,\omega},
\end{equation}
with zero boundary condition $\phi\vert_{traj}=0$ on the mirror's trajectory.
Here the variables $u=t-x,\,v=t+x$ are used and the mirror's (or charge's)
trajectory on the $u,\,v$ plane is given by any of the two mutually inverse
functions $v^{mir}=f(u),\;u^{mir}=g(v)$. 

For the $in$- and $out$-sets of massless Dirac equation solutions see [3]. Dirac
solutions differ from (1) by the presence of bispinor coefficients at $u$- and
$v$-plane waves. The current densities corresponding to these solutions have 
only tangential components on the boundary. So, the boundary condition both for
scalar and spinor field is purely geometrical, it does not contain any 
dimentional parameters.

The Bogoliubov coefficients $\alpha_{\omega'\omega},\,\beta_{\omega'\,\omega}$
appear as the coefficients of the expansion of the solutions of the $out$-set
in the solutions of the $in$-set and $\alpha^*_{\omega'\omega},\,\mp\beta_
{\omega'\omega}$ as the coefficients of the inverse expansion. The upper and
lower signs correspond to scalar (Bose) and spinor (Fermi) field. Then the mean
number of quanta with frequency $\omega$ and wave vector $\omega>0$ radiated by
accelerated mirror to the right semispace is given by the integral
\begin{equation}      
d\bar n_{\omega}=\frac{d\omega}{2\pi}\int_0^\infty \frac{d\omega'}{2\pi}
\vert \beta_{\omega'\omega}\vert^2.
\end{equation}

At the same time the spectra of photons and scalar quanta emitted by electric
and scalar charges moving along the trajectory $x_{\alpha}(\tau)$ in 3+1 space
are defined by the Fourier transforms of the electric current density 4-vector
$j_{\alpha}(k)$ and the scalar charge density $\rho(k)$,
\begin{equation}
s=1,\quad j_{\alpha}(k)=e\int d\tau\,\dot x_{\alpha}(\tau)\,e^{-ik^{\alpha}
x_{\alpha}(\tau)},
\end{equation}
\begin{equation}
d\bar n^{(1)}_k=\vert j_{\alpha}(k_+,k_-)\vert ^2\frac{dk_+dk_-}{(4\pi)^2},
\end{equation}
\begin{equation}
s=0,\quad \rho (k)=e\int d\tau\,e^{-ik^{\alpha}x_{\alpha}(\tau)},
\end{equation}
\begin{equation}
d\bar n^{(0)}_k=\vert \rho (k_+,k_-)\vert ^2\frac{dk_+dk_-}{(4\pi)^2},
\end{equation}
where $s$ and $k^{\alpha}$ are the spin and 4-momentum of quanta,
$$
k^2=k^2_1+k^2_{\perp}-k^2_0=0,\quad k^2_{\perp}=k^2_0-k^2_1=k_+k_-,\quad
k_{\pm}=k^0\pm k^1,
$$
and in (4) and (6) it is supposed that the trajectory $x^{\alpha}(\tau)$ has
only $x^0$ and $x^1$ nontrivial components.

The symmetry between the creation of Bose or Fermi pairs by accelerated mirror
in 1+1 space and the emission of single photons or scalar quanta by electric or
scalar charge in 3+1 space consists, first of all, in the coincidence of the
spectra. If one puts $2\omega=k_+,\;2\omega'=k_-$, then
\begin{equation}
\vert\beta^B_{\omega'\omega}\vert^2=\frac{1}{e^2}\vert j_{\alpha}(k_+,k_-)\vert
^2,\qquad \vert\beta^F_{\omega'\omega}\vert^2=\frac{1}{e^2}\vert\rho(k_+,k_-)
\vert^2.
\end{equation}
More refined assertion for the Bose case:
\begin{equation}
\beta^{B*}_{\omega'\omega}=-\sqrt{\frac{k_+}{k_-}}\frac{j_-(k)}{e}=
\sqrt{\frac{k_-}{k_+}}\frac{j_+(k)}{e}=\frac{\varepsilon_{\alpha\beta}k^{\alpha}
j^{\beta}(k)}{e\sqrt{k_+k_-}},
\end{equation}
\begin{equation}
j_-(k)=e\int du\,e^{\frac i2(k_+u+k_-f(u))},\quad
j_+(k)=e\int dv\,e^{\frac i2(k_-v+k_+g(v))}.
\end{equation}
The 2-vectors $j_{\alpha}(k)$ and $a_{\beta}(k)=\varepsilon_{\alpha\beta}
k^{\alpha}/\sqrt{k_+k_-}$ are spacelike for timelike $k^{\alpha}$ and in a
system where $k_+=k_-$ or $\omega=\omega'$ they have only spatial components
preciesely equal to $e\beta^{B*}_{\omega'\omega}$ and $1$ correspondingly.

And for the Fermi case:
\begin{equation}
\beta^{F*}_{\omega'\omega}=\frac {1}{e}\rho (k).
\end{equation}

In the present paper in Section 2 we underline the symmetry of analytical 
expressions for the Bogoliubov coefficients $\alpha ,\,\beta^*$ and, at the same 
time, the physical distinction between them: $\beta^{B,F*}$ is the amplitude
of a source of the waves which are bilinear in massless Bose or Fermi fields 
and carry the timelike momenta, whereas $\alpha^{B,F}$ is the amplitude of a 
source of the similar waves but which carry the spacelike momenta, see (14), 
(15). In Sections 3 and 4 it is shown that the waves with timelike momenta 
emitted and absorbed by the source are involved in the forming of the imaginery 
part of the source selfaction. This physical picture is naturally embodied in 
the integral relation (20) between propagators $\Delta_2 (z,m)$ of virtual pairs 
with masses $m,\:\mu\leqslant m <\infty$, in two-dimentional spacetime and 
propagator $\Delta_4 (z,\mu)$ of the particle in four-dimentional spacetime. The 
analytical properties of the expressions appeared allow to define too the real 
part of the selfaction. This leads to coincidence of the selfactions and, hence, 
of the vacuum-vacuum amplitudes of the mirror and the charge if one considers 
that $e^2=1$. In Section 5 the fields of perturbations carring the spacelike
momenta are considered. These fields are defined by the matrixes $\alpha^{B,F}$.
Their Lorentz-invariant traces $\pm {\rm tr}\,\alpha^{B,F}$ are considered in
Section 6. They describe correspondingly the vector and scalar interactions of
the accelerated mirror with the uniformly moving detector in the neighbourhood
of the point of contact of their trajectories. In Sections 7 and 8 the traces
$\pm {\rm tr}\,\alpha^{B,F}$ are found for the three concrete trajectories   
permitting the analytical solutions. At the same place the general expressions 
for the traces are given and their ultraviolet and infrared singularities are 
considered. In these Sections the comparison of the found traces $\pm {\rm tr}
\,\alpha^{B,F}$ with the mass shifts $\Delta m_{1,0}$ of the electric and scalar 
charges, moving along the same trajectory as the mirror's one but in 3+1-space,
is carried out. In this connection, in Sector 9 the mass shifts $\Delta m_{1,0}$
of the charges moving along the exponential trajectory are found. In Concludions
we discuss the connection of the traces $\pm {\rm tr}\,\alpha^{B,F}$ with the 
general definition of the selfaction accounting for the interference effects, 
and call attention to the fact that the symmetry fixes the value of the bare
charge squared, $e^2_0=1$, that corresponds to bare fine structure constant
$\alpha_0=1/4\pi$. The smallness and geometrical origin of this value may be
interesting in quantum electrodynamics. In Appendix the even singular solutions
of inhomogeneous wave equations with mass and momentum transfer parameters are
considered. The integral relations (20) and (100) between these solutions for
1+1- and 3+1-spaces are very important for the symmetry considered.

\section{The physical interpretation of $\beta^{*}_{\omega'\omega}$}

The absolute pair production amplitude and single-particle scattering amplitude
are connected by the relation [4]
\begin{equation}
\langle {\rm out}\omega''\omega\vert {\rm in}\rangle=-\sum_{\omega'}
\langle {\rm out}\omega''\vert \omega'{\rm in}\rangle\,\beta^*_{\omega'\omega}.
\end{equation}
The $\beta^*_{\omega'\omega}$ was interpreted as the amplitude of a source of
a pair of the massless particles potentially emitted to the right and to the    
left with frequences $\omega$ and $\omega'$ respectively. While the particle
with frequency $\omega$ actually escaped to the right the particle with
frequency $\omega'$ propagates some time then is reflected by the mirror and
is actually emitted to the right with altered frequency $\omega''$, see Fig. 1.

On the time interval between pair creation and reflection of the left particle
we have the virtual pair with energy $k^0$, momentum $k^1$, and mass $m$:
\begin{equation}
k^0=\omega+\omega',\quad k^1=\omega-\omega',\quad m=\sqrt{-k^2}=2\sqrt{\omega
\omega'}.
\end{equation}

Apart from the polar timelike 2-vector $k^{\alpha}$, very important is the axial
spacelike 2-vector $q^{\alpha}$,
\begin{equation}
q_{\alpha}=\varepsilon_{\alpha\beta}k^{\beta},\quad q^0=-k^1=-\omega+\omega',
\quad q^1=-k^0=-\omega-\omega'<0.
\end{equation}
With the help of $k^{\alpha}$ and $q^{\alpha}$ the symmetry between $\alpha$
and $\beta$ coefficients becomes clearly expressed:
\begin{equation}
s=1,\qquad e\beta^{B*}_{\omega'\omega}=-\frac{q_{\alpha}j^{\alpha}(k)}
{\sqrt{k_+k_-}},\qquad e\alpha^B_{\omega'\omega}=-\frac{k_{\alpha}j^{\alpha}(q)}
{\sqrt{k_+k_-}},
\end{equation}
\begin{equation}
s=0,\qquad e\beta^{F*}_{\omega'\omega}=\rho(k),\qquad e\alpha^F_{\omega'\omega}
=\rho(q).
\end{equation}

Note that the equations (3) and (5) define the current density $j^{\alpha}(k)$
and the charge density $\rho (k)$ as the functionals of the trajectory 
$x^{\alpha}(\tau)$ and the functions of any 2- or 4-vector $k^{\alpha}$. It can
be shown that in 1+1-space $j^{\alpha}(k)$ and $j^{\alpha}(q)$ are the spacelike 
and timelike polar vectors correspondingly if $k^{\alpha}$ and $q^{\alpha}$ 
are the timelike and spacelike vectors.

The boundary condition on the mirror evokes in the vacuum of massless scalar or
spinor field the appearance of vector or scalar disturbance waves bilinear in
massless fields. There are two types of these waves:

1) The waves with amplitude $\alpha_{\omega'\omega}\;(\alpha^*_{\omega'\omega})$
which carry the spacelike momentum directed to the left (right), and

2) The waves with amplitude $\beta^*_{\omega'\omega}\;(\beta_{\omega'\omega})$
which carry the timelike momentum with positive (negative) frequency.

The waves with the spacelike momenta appear even if the mirror is in rest or
moves uniformly (Casimir effect), while the waves with the timelike momenta
appear only in the case of accelerated mirror.

The pair of Bose (Fermi) particles has spin 1 (0) because its source is the
current density vector (charge density scalar), see [5] or the problem 12.15 
in [6].

\section{The appearance of the mass in massless theory and of the
invariant singular solutions of wave equation with mass}

The bilinear in massless bose-field disturbances, defined by the amplitudes
$\beta^{B*}_{\omega'\omega}$, form, according to (8), the positive-frequency
current density vector. Its minus-component in the point $U,V$
can be represented as
\begin{eqnarray}
\int\!\!\!\!\int\limits_0^\infty\frac{d\omega d\omega'}{(2\pi)^2}\,\frac{1}{e}j_-(k)
\,e^{-i\omega U-i\omega' V}=
\frac{1}{8\pi^2}\int du\int\limits_0^\infty d\rho \rho
\int\limits_{-\infty}^\infty d\theta\,e^{-i\rho(z^0{\rm ch}\theta-
z^1{\rm sh}\theta)},
\end{eqnarray}
if instead of $\omega,\,\omega'$ the hyperbolic variables $\rho,\,\theta$ are
used,
\begin{equation}
d\omega d\omega'=\frac12\rho d\rho d\theta,\quad \omega=\frac12\rho e^{\theta},
\quad \omega'=\frac12\rho e^{-\theta},\quad \rho=2\sqrt{\omega\omega'},
\quad \theta=\ln\sqrt{\frac{\omega}{\omega'}},
\end{equation}
and $z^{\alpha}=x^{\alpha}-x^{\alpha}(\tau)$, see Fig. 2.

As it is seen from (12) $\rho=m$ is the mass of the pair and $\theta$ is the
rapidity. The integral over rapidity in (16) is the well known invariant
positive-frequency singular function of wave equation for 2-dimentional spacetime:
\begin{eqnarray}
\int_{-\infty}^\infty d\theta\,e^{-im(z^0{\rm ch}\theta-z^1{\rm sh}\theta)}=
-4\pi i\Delta^+_2(z,m)=    \nonumber\\
2\theta(-z^2)K_0(i\varepsilon(z^0)m\sqrt{-z^2})+2\theta(z^2)K_0(m\sqrt{z^2}),
\end{eqnarray}
\begin{equation}
(\partial^2_t-\partial^2_x+m^2)\Delta^+_2(z,m)=0.
\end{equation}
This function describes the wave field of pairs with mass $m$ and any possible
positive-frequency momenta. According to it the pairs are created, propagated
and absorbed near the mirror within spacelike interval of the order of $m^{-1}$.

By using the very important integral relation between the singular functions of
wave equations for $d$- and $d+2$-dimentional spacetimes,
\begin{equation}
\Delta^f_{d+2}(z,\mu)=\frac{1}{4\pi}\int_{\mu^2}^\infty dm^2\,\Delta^f_d (z,m),
\end{equation}
the right-hand side of (16) can be represented in the form
\begin{equation}
-\frac{i}{4\pi}\int du\int\limits_{\mu^2\to 0}^\infty dm^2\,\Delta^+_2(z,m)=
-i\int\limits_{-\infty}^\infty du\,\Delta^+_4(z,\mu).
\end{equation}
The small mass $\mu$ is retained to eliminate the infrared divergence in the
following.

Analogously, the positive-frequency plus-component of the current density in the
point $U,V$ can be represented as
\begin{equation}
\int\!\!\!\!\int\limits_0^\infty\frac{d\omega d\omega'}{(2\pi)^2}\,\frac{1}{e}j_+(k)
\,e^{-i\omega U-i\omega' V}=-i\int\limits_{-\infty}^\infty dv\,\Delta^+_4(z,\mu).
\end{equation}
The differentials $du$ and $dv$ in (21) and (22) may be replaced by $d\tau\,
\dot x_-(\tau)$ and $d\tau\,\dot x_+(\tau)$.

The bilinear in massless fermi-field disturbances defined by the amplitudes
$\beta^{F*}_{\omega'\omega}$ form the positive-frequency charge density scalar.
At the point $U,V$ it can be represented by
\begin{equation}
\int\!\!\!\!\int\limits_0^\infty\frac{d\omega d\omega'}{(2\pi)^2}\,\frac{1}{e}
\rho(k)\,e^{-i\omega U-i\omega' V}=-i\int\limits_{-\infty}^\infty d\tau\,
\Delta^+_4(z,\mu).
\end{equation}
If we put the point $U,V$ on the trajectory, so that $U=x_-(\tau'),\,
V=x_+(\tau')$ and $z^{\alpha}=x^{\alpha}(\tau')-x^{\alpha}(\tau)$, and integrate
(21) over $V$ and (22) over $U$ then their half of the sum will differ from
the tr$\,\beta^+\beta$ only by the multiplier $i$:
\begin{eqnarray}
{\rm tr}\,\beta^{B+}\beta^B\equiv\int\!\!\!\!\int\limits_0^\infty
\frac{d\omega d\omega'}{(2\pi)^2}\vert\beta^B_{\omega'\omega}\vert^2= \nonumber\\
\frac{i}{2}\int\!\!\!\!\int(du\,dV+dv\,dU)\,\Delta^+_4(z,\mu)=
-i\int\!\!\!\!\int d\tau d\tau'\,\dot x_{\alpha}(\tau)\dot x^{\alpha}
(\tau')\,\Delta^+_4(z,\mu).
\end{eqnarray}
As the function $\Delta^+$ has odd in $z$ real part and even in $z$ imaginery
part which are connected with the even in $z$ causal (Feynman) function
$\Delta^f$,
\begin{equation}
\Delta^+(z,\mu)=\frac12\Delta(z,\mu)+\frac{i}{2}\Delta^1(z,\mu),\quad
{\rm Re}\,\Delta^+=\varepsilon(z^0){\rm Re}\,\Delta^f,\quad {\rm Im}\,\Delta^+=
{\rm Im}\,\Delta^f,
\end{equation}
the tr$\beta^{B+}\beta^B$ may be written in the form
\begin{equation}
{\rm tr}\,(\beta^+\beta)^B={\rm Im}\,\int\!\!\!\!\int d\tau d\tau'
\dot x_{\alpha}(\tau)\dot x^{\alpha}(\tau')\,\Delta^f_4(z,\mu).
\end{equation}
The ${\rm tr}\,\beta^{F+}\beta^F$ can be obtained from the right-hand side of 
(26) by the substitution $\dot x_{\alpha}(\tau)\dot x^{\alpha}(\tau')\,\to\,1$.

\section{Vacuum-vacuum amplitude $\langle {\rm out}\vert {\rm in}\rangle =e^{iW}$}
According to DeWitt [7], Wald [8] and others (including myself [4])
\begin{equation}
2\,{\rm Im}\,W^{B,F}=\pm\frac12{\rm tr}\,\ln(1\pm\beta^+\beta)\quad{\rm or}\quad
\pm{\rm tr}\,\ln(1\pm\beta^+\beta)
\end{equation}
correspondingly to the cases when particle is identical or nonidentical to
antiparticle. We confine ourselves by the last case and tr$\,\beta^+\beta\ll 1$. Then
\begin{equation}
2\,{\rm Im}\,W^{B,F}={\rm Im}\int\!\!\!\!\int d\tau d\tau'\left\{
\begin{array}{c}
\dot x_{\alpha}(\tau)\dot x^{\alpha}(\tau')\\1
\end{array}
\right\}\Delta^f_4(z,\mu).
\end{equation}

We may omit the Im-signs from both of sides of this equation and define the
actions for bose- and fermi-mirrors in 1+1-space as
\begin{equation}
W^{B,F}=\frac12\int\!\!\!\!\int d\tau d\tau'\left\{
\begin{array}{c}
\dot x_{\alpha}(\tau)\dot x^{\alpha}(\tau')\\1
\end{array}
\right\}\Delta^f_4(z,\mu).
\end{equation}
Compare this with the well known actions for electric and scalar charges in
3+1-space:
\begin{equation}
W^{1,0}=\frac12\,e^2\int\!\!\!\!\int d\tau d\tau'\left\{
\begin{array}{c}
\dot x_{\alpha}(\tau)\dot x^{\alpha}(\tau')\\1
\end{array}
\right\}\Delta^f_4(z,\mu).
\end{equation}
The symmetry would be complete if $e^2=1$, i.e. if the fine structure constant
were $\alpha=1/4\pi$. This "ideal" value of fine structure constant for the
charges would correspond to the ideal, geometrical boundary condition on the 
mirror.

For the mirror trajectory with nonzero relative velocity $\beta_{21}$ of its
ends (nonzero relative rapidity $\theta={\rm Arth}\,\beta_{21}$) the changes
of actions due to acceleration are given by the formulae
\begin{equation}
{\rm Re}\,\Delta W^B=\frac1{8\pi}(\frac{\theta}{{\rm th}\,\theta}-1),\qquad
{\rm Re}\,\Delta W^F=\frac1{8\pi}(1-\frac{\theta}{{\rm sh}\,\theta}).
\end{equation}
For uniformly accelerated mirror with proper acceleration $a$ its
velocity $\beta(\tau)={\rm th}\,a\tau,\;\tau$ is the proper time. Then
$\theta=a(\tau_2-\tau_1)$ and for $\tau_2-\tau_1\,\to\,\infty$
\begin{equation}
{\rm Re}\,\Delta W^B=\frac{\vert a\vert}{8\pi}\,(\tau_2-\tau_1).
\end{equation}
By definition, the
\begin{equation}
{\rm Re}\,\Delta m^B=-\frac{\partial {\rm Re}\,\Delta W^B}{\partial \tau_2}=
-\frac{\vert a\vert}{8\pi}
\end{equation}
is the self-energy shift of bose-mirror at acceleration. It differs from the
mass shift of uniformly accelerated electron only by the absence of the
multiplier $e^2=4\pi\alpha$. The self-energy shift of uniformly accelerated
fermi-mirror ${\rm Re}\,\Delta m^F=0$.

There are two arguments in favour of definition of action by means of the causal
function $\Delta^f_4(z,\mu)$.

1. The action must represent not only the radiation of real quanta but also the
self-energy and polarization effects. While the first effects are described by
the solutions of homogeneous wave equation the second ones require the
inhomogeneous wave equation solutions which contain information about proper
field of a source. Namely such solutions of homogeneous and inhomogeneous wave
equations are the functions $(1/2)\Delta^1={\rm Im}\,\Delta^f$ and
$\bar \Delta={\rm Re}\,\Delta^f$.

2. While the appearance of $(1/2)\Delta^1\equiv{\rm Im}\,\Delta^f$ in the
imaginary part of the action is the consequence of the mathematical
transformations of the integral
$$
\int\!\!\!\!\int\limits_0^\infty\frac{d\omega d\omega'}{(2\pi)^2}
\vert\beta_{\omega'\omega}\vert^2,
$$
transformations similar to the Plancherel theorem, the function $\bar \Delta
\equiv{\rm Re}\,\Delta^f$ in the real part of the action will be unique if it
appears as the real part of that analytical continuation of the function
$(i/2)\Delta^1(z,\mu)$ to the negative $z^2$ which is even in $z$ as $\Delta^1$
itself.

In conclusion of the third and fourth sections it should be noted that due to
the transition from variables $\omega,\,\omega'$ to hyperbolic variables $\rho,
\,\theta$, which reflect the Lorentzian symmetry of the problem, both the
function $\Delta_2(z,m)$, describing the propagation of virtual pair with mass
$m=\rho=2\sqrt{\omega\omega'}$ in two-dimentional space-time, and the mass
spectrum of these pairs arise. Further integration over the mass leads to the
function $\Delta_4(z,\mu)$, which coincides with the propagator of a particle
moving in four-dimentional space-time with the mass $\mu$ equal to the least
mass of virtual pairs. Thus, the relation (20) appears in the framework of the
present method and is immanent to the symmetry, connecting the processes in
two- and four-dimentional space-times.

In the paper [9] on the same topic as this one the relation (20) was obtained
by the author as independent of the processes considered and was in need
of proof that the integration variable figuring in it coincides with the pair
mass $m=2\sqrt{\omega\omega'}$ indeed.

\section{Formation of tachyon disturbances with invariant momentum transfer}
The bilinear in massless bose-field perturbations, which are defined by the
amplitudes $\alpha^B_{\omega'\omega}$ and carry to the left the spacelike
momenta, can be represented in the point $U,V$ by the two current density
components
\begin{eqnarray}
\int\!\!\!\!\int\limits_0^\infty\frac{d\omega d\omega'}{(2\pi)^2}\,\frac{1}{e}
j_{\pm}(q)\,e^{i\omega U-i\omega' V}=
\frac{1}{8\pi^2}\int d\tau\,\dot x_{\pm}(\tau)\int\limits_0^\infty d\rho\rho\int\limits_{-\infty}
^\infty d\theta\,e^{i\rho(z^0{\rm sh}\theta-z^1{\rm ch}\theta)}
\end{eqnarray}
if one again uses the change of variables (17) and the notation $z^{\alpha}=
x^{\alpha}-x^{\alpha}(\tau)$.
Now the integral over $\theta$ is equal to
\begin{eqnarray}
\int_{-\infty}^\infty d\theta\,e^{i\rho(z^0{\rm sh}\theta-z^1{\rm ch}\theta)}=
4\pi i\,\Delta^L_2(z,\rho)=   \nonumber\\
2\theta(-z^2)K_0(\rho\sqrt{-z^2})+
2\theta(z^2)K_0(i\varepsilon(z^1)\rho\sqrt{z^2}).
\end{eqnarray}
The integrand in the left-hand side of (35) is the wave with spacelike 2-momentum
$q^{\alpha}:\quad q^1=-\omega-\omega'=-\rho\,{\rm ch}\theta,\quad q^0=-\omega+
\omega'=-\rho\,{\rm sh}\theta,\quad \rho=\sqrt{q^2}$. The function $\Delta^L_2
(z,\rho)$ is the superposition of plane waves with spacelike momenta
directed to the left and with fixed invariant momentum transfer $\rho=2\sqrt
{\omega\omega'}$. It satisfies the wave equation with negative mass squared:
\begin{equation}
(\partial^2_t-\partial^2_x-\rho^2)\Delta^L_2(z,\rho)=0.
\end{equation}
By using the integral relation analogous to relation (20) (see Appendix) the 
right-hand side of (34) can be represented in the form
\begin{equation}
\frac{i}{4\pi}\int d\tau\,\dot x_{\pm}(\tau)\int\limits_{\nu^2\to 0}^\infty
d\rho^2\,\Delta^L_2(z,\rho)=-i\int d\tau\,\dot x_{\pm}(\tau)\,\Delta^L_4(z,\nu).
\end{equation}
The small momentum transfer $\nu$ is retained to eliminate the infrared 
divergence in what follows.

Analogously, the bilinear in fermi-field disturbances, which are defined by the
amplitudes $\alpha^F_{\omega'\omega}$ and carry the left-directed spacelike
momenta, form the charge density scalar. It can be represented in the point
$U,V$ by the integral
\begin{equation}
\int\!\!\!\!\int\limits_0^\infty\frac{d\omega d\omega'}{(2\pi)^2}\,\frac{1}{e}
\rho (q)\,e^{i\omega U-i\omega' V}=-i\int d\tau\,\Delta^L_4(z,\nu).
\end{equation}

These representations can be useful in the problems close to static ones where
apart from acceleration or instead of it another characteristic length enters.

\section{Interpretation of the traces $\pm{\rm tr}\,\alpha^{B,F}$ of Bogoliubov
coefficients}
The invariant description of the mirror's trajectory on the $u,v$-plane demands
that the function $u^{mir}=g(v)$ contains the two positive parameters $\varkappa,
\varkappa'$, transforming as $x_+=v,\,x_-=u,$ and actually connects the invariant
variables $\varkappa u,\,\varkappa' v$ between themselves:
\begin{equation}
u^{mir}=g(v)=\frac 1 \varkappa\,G(\varkappa'v).
\end{equation}
Its expansion near the origin of coordinates $u=v=0$, being on the trajectory,
has the form
\begin{equation}
g(v)=\frac 1 \varkappa (\varkappa'v+b\varkappa'^2v^2+\frac13\,c\varkappa'^3v^3+
\ldots ),
\end{equation}
where $b,c,\ldots\,$-some numbers. Since the mirror's velocity $\beta (v)$ and
proper acceleration $a(v)$ are defined by the formulae
\begin{equation}
\beta(v)=\frac{1-g'(v)}{1+g'(v)},\qquad a(v)=-\frac{g''(v)}{2g'^{3/2}(v)},
\end{equation}
the two first coefficients of the expansion (40) define the mirror's velocity
$\beta_0$ and acceleration $a_0$ at zero point:
\begin{equation}
\beta_0=\frac{1-\varkappa'/\varkappa}{1+\varkappa'/\varkappa},\qquad
a_0=-b\sqrt{\varkappa\,\varkappa'}.
\end{equation}
The absolute value of acceleration at zero point will be denoted as $w_0=
\vert b\vert\sqrt{\varkappa\,\varkappa'}$.

Let us define the Lorentz-invariant ${\rm tr}\,\alpha$ by the formula
\begin{equation}
{\rm tr}\,\alpha=\int\!\!\!\!\int\limits_0^\infty\frac{d\omega d\omega'}
{(2\pi)^2}\,\alpha_{\omega' \omega}\,2\pi\,\delta\left(\sqrt{\frac{\varkappa'}
{\varkappa}}\omega-\sqrt{\frac{\varkappa}{\varkappa'}}\omega'\right),
\end{equation}
in which the Lorentz-invariant argument of $\delta$-function is the difference
of frequences
\begin{equation}
\Omega=\sqrt{\frac{\varkappa'}{\varkappa}}\omega,\qquad
\Omega'=\sqrt{\frac{\varkappa}{\varkappa'}}\omega'
\end{equation}
of reflected and incident waves in the proper system of the mirror at the moment
$u=v=0$. According to (42) the multipliers $\sqrt{\varkappa'/\varkappa},\,
\sqrt{\varkappa/\varkappa'}$ figuring in formula (44) are the Doppler factors
connecting the frequences in the laboratory and proper systems. In proper system
of the mirror $\Omega=\Omega'=\sqrt{\omega\,\omega'}$.

According to (43) the ${\rm tr}\,\alpha$ is Lorentz-invariant and dimentionless
quantity or, perhaps, has dimentionality of action, since $\hbar=1$. Let us
consider its physical meaning. For this let us turn to equality of expressions
(34) and (37)
\begin{eqnarray}
\int\!\!\!\!\int\limits_0^\infty\frac{d\omega d\omega'}{(2\pi)^2}\,\frac{1}{e}
j_{\pm}(q)\,e^{i\omega U-i\omega' V}=
-i\int d\tau\,\dot x_{\pm}(\tau)\,\Delta^L_4(z,\nu),
\end{eqnarray}
where $z^\alpha=x^\alpha-x^\alpha (\tau),\; x_-=U,\:x_+=V.$ Let us put the point
$U,\,V$ on the tangent line to the mirror's trajectory at zero point, so that
\begin{equation}
U=X_-(\tau')=\sqrt{\frac{\varkappa'}{\varkappa}}\,\tau',\quad
V=X_+(\tau')=\sqrt{\frac{\varkappa}{\varkappa'}}\,\tau'
\end{equation}
$\tau'$-is the proper time of the point on the tangent line, and integrate the
both sides of equation (45) over $dU=\dot X_-\,d\tau'$ or $dV=\dot X_+\,d\tau'$
correspondingly the upper or lower sign in (45). Then accounting for the formula
(14) and current conservation we obtain on the left the ${\rm tr}\,\alpha$
both for upper and lower signs in (45). On the right we obtain the integral
\begin{equation}
-i\int\!\!\!\!\int\,d\tau\,d\tau'\,\dot x_{\pm}(\tau)\dot X_{\mp}(\tau')\,
\Delta_4^L(z,\nu),\qquad z^\alpha=X^\alpha (\tau')-x^\alpha (\tau),
\end{equation}
in which according to the result for the left part one may left instead of
\begin{equation}
\dot x_{\pm}(\tau)\dot X_{\mp}(\tau')=-\dot x_\alpha (\tau)\dot X^\alpha (\tau')
\mp \varepsilon_{\alpha\beta}\,\dot x^\alpha (\tau)\dot X^\beta (\tau')
\end{equation}
only the first term symmetrical with respect to permutation $\dot x_\alpha (\tau)
\rightleftarrows \dot X_\alpha (\tau')$. Thus we obtain

\begin{equation}
{\rm tr}\,\alpha^B=i\int\!\!\!\!\int\,d\tau\,d\tau'\,\dot x_\alpha (\tau)
\dot X^\alpha (\tau')\,\Delta_4^L(z,\nu),\qquad 
z^\alpha=X^\alpha (\tau')-x^\alpha (\tau).      
\end{equation}

Integrating similarly the both parts of equation (38) along tangent line (46) 
and taking into account the formulae (15) and (43) we obtain
\begin{equation}
{\rm tr}\,\alpha^F=-i\int\!\!\!\!\int\,d\tau\,d\tau'\,\Delta_4^L(z,\nu),\qquad 
z^\alpha=X^\alpha (\tau')-x^\alpha (\tau).
\end{equation}

For the trajectories situated on the Minkowsky plane on the left from their
tangent line at zero point the coordinate $z^1\geqslant 0$. In this case the
$\Delta_4^L(z,\nu)$ can be replaced by the function
\begin{equation}
\Delta_4^{LR}(z,\nu)=\frac{1}{4\pi}\delta(z^2)-\frac{\nu}{8\pi\sqrt{z^2}}
\theta(z^2)[J_1(\nu \sqrt{z^2})-iN_1(\nu \sqrt{z^2})]+i\frac{\nu}{4\pi^2\sqrt
{-z^2}}\theta(-z^2)K_1(\nu\sqrt{-z^2}),
\end{equation}
which differs from the causal function $\Delta_4^f (z,\mu)$ by complex 
conjugation and replacement $\mu\to i\nu$ (or by the replacement $z^2\to-z^2,\:
\mu\to\nu$). More detail about this function see in Appendix.

Thus for the mentioned trajectories
\begin{equation}
\pm{\rm tr}\,\alpha^{B,F}=i\int\!\!\!\!\int d\tau d\tau'\left\{
\begin{array}{c}
\dot x_{\alpha}(\tau)\dot X^{\alpha}(\tau')\\1
\end{array}
\right\}\Delta^{LR}_4(z,\nu),\qquad z^\alpha=X^\alpha (\tau')-x^\alpha (\tau).
\end{equation}
The expression obtained allows to interpret $\pm{\rm tr}\alpha^{B,F}$ as a
functional describing the interaction of two vector or scalar sources by means 
of exchange by vector or scalar quanta with spacelike momenta. At the same time
one of the sources moves along the mirror's trajectory while another one moves
along the tangent line to it at zero point. The last source can be considered
as a probe or detector of exitation created by the accelerated mirror in vacuum.

\section{The traces of Bogoliubov's coefficients for hyperbolic and exponential
trajectories}
Let us consider the ${\rm tr}\,\alpha^{B,F}$ for hyperbolic mirror's trajectory
\begin{equation}
u^{mir}=g(v)=\frac{\varkappa' v}{\varkappa (1-\varkappa' v)}.
\end{equation}
Using the formulae (14) and (4) of the paper [3] it is not difficult to represent
$\alpha^{B,F}_{\omega'\omega}$ via Macdonald's functions $K_{1,0}$:
\begin{equation}
\alpha^{B,F}_{\omega'\omega}=\frac{2}{\sqrt{\varkappa \varkappa'}}\,e^{i(
\frac{\omega}{\varkappa}+\frac{\omega'}{\varkappa'})}\,
K_{1,0}\left(2i\sqrt{\frac{\omega\omega'}{\varkappa\varkappa'}}\right). 
\end{equation}
Then according to the formula (43)
\begin{equation}
{\rm tr}\,\alpha^{B,F}=\frac1\pi\,\int_0^\infty d\left(\frac\omega\varkappa\right)
\,e^{2i\frac{\omega}{\varkappa}}\,K_{1,0}\left(2i\frac{\omega}{\varkappa}
\right)=\frac{1}{2\pi}\,\int_0^\infty dz\,e^{iz}\,K_{1,0}(iz).
\end{equation}
The variable $z$ in this integral has a simple physical meaning: it is equal to 
the ratio of invariant momentum transfer to invariant proper acceleration at
zero point (but for hyperbolic motion the acceleration is the same on the whole 
trajectory):
\begin{equation}
z=\frac{\rho}{w_0},\qquad\rho=2\sqrt{\omega\omega'},\qquad w_0=\sqrt{\varkappa
\varkappa'}.
\end{equation}

The ultraviolet divergency of the integral (55) is removed by subtraction from
the integrand its asymptotics for $z\to\infty$. The infrared divergency (for the 
Bose-case) is removed by introducing the nonzero lower limit $\varepsilon=
\nu /w_0 \ll 1$, defined by the minimal momentum transfer $\nu$. As a result we
obtain the integral
\begin{equation}
{\rm tr}\,\alpha^{B,F}=\frac{1}{2\pi}\int\limits_{s\varepsilon}^\infty dz\,
[e^{iz}\,K_s(iz)-\sqrt{\frac{\pi}{2iz}}],\quad s=1,\,0,\quad\varepsilon\ll 1.
\end{equation}
Now the integration contour can be turned on the negative imaginery semiaxis
going round (in Bose-case) the singularity at zero along the arc of a circle
with small radius $\varepsilon$. The further calculation leads to the simple
expressions
\begin{equation}
{\rm tr}\,\alpha^B=\frac{1}{2\pi}[-\frac\pi 2-i\left(\ln\frac{2w_0}{\gamma\nu}
-1\right)],\qquad\nu\ll w_0,\quad\gamma=1,781\ldots ,
\end{equation}
\begin{equation}
{\rm tr}\,\alpha^F=\frac{1}{2\pi}i.
\end{equation}

For the exponentional mirror's motion
\begin{equation}
u^{mir}=-\frac{1}{\varkappa}\ln (1-\varkappa'v),\qquad 
v^{mir}=\frac{1}{\varkappa'}-\frac{1}{\varkappa'}e^{-\varkappa u},
\end{equation}
the same formulae (14) and (4) from [3] leads to the Bogoliubov's coefficients
\begin{equation}
\alpha^B_{\omega'\omega}=\frac1\varkappa\sqrt{\frac{\omega}{\omega'}}\Gamma
\left(\frac{i\omega}{\varkappa}\right)\,e^{i\frac{\omega'}{\varkappa'}-\frac
{i\omega}{\varkappa}\ln\frac{i\omega'}{\varkappa'}},
\end{equation}
\begin{equation}
\alpha^F_{\omega'\omega}=\frac{1}{\sqrt{i\varkappa\omega'}}\Gamma \left(\frac12+
\frac{i\omega}{\varkappa}\right)\,e^{\frac{i\omega'}{\varkappa'}-\frac{i\omega}
{\varkappa}\ln\frac{i\omega'}{\varkappa'}}.
\end{equation}
The traces ${\rm tr}\,\alpha^{B,F}$ which divergences were removed by the above 
mentioned prescription are such
\begin{equation}
{\rm tr}\,\alpha^B=\frac{1}{2\pi}\int_{\varepsilon}^\infty dx\,[\Gamma (ix)
e^{ix-ix\ln ix}-\sqrt{\frac{2\pi}{ix}}],
\end{equation}                          
\begin{equation}
{\rm tr}\,\alpha^F=\frac{1}{2\pi}\int_0^\infty dx\,[\Gamma (\frac12+ix)\frac
{e^{ix-ix\ln ix}}{\sqrt{ix}}-\sqrt{\frac{2\pi}{ix}}].      
\end{equation}
In these integrals the variable $x$ is equal to one fourth of the variable $z$
which has, as well as in (56), the sense of momentum transfer in units of $w_0$:
\begin{equation}
x=\frac14 z,\quad z=\frac{\rho}{w_0},\quad \rho=2\sqrt{\omega \omega'},\quad
w_0=\frac12\sqrt{\varkappa \varkappa'}.
\end{equation}
Analogously, $\varepsilon=\nu /4w_0 \ll 1$. Note that at exponential motion (60)
the proper acceleration increases from zero to infinity and as a function of
proper time $\tau$ is given by the formula 
\begin{equation}
a(\tau)=-\frac{w_0}{1-w_0\tau}.
\end{equation}

Now it is not difficult to see that the subtractive terms in integrals (63),(64)
exactly coinside with the similar terms in integrals (57) if one expreses them 
via physical variable $z$. In other words, up to removing ultraviolet divergency
from the integrals defining ${\rm tr\,\alpha}$ the asymptotical behaviour of
the integrands in the variable $z=\rho/w_0 \to \infty$ is described by the
universal formula
\begin{equation}
\frac{1}{2\pi}\sqrt{\frac{\pi}{2iz}}.
\end{equation}
It will be shown in the next section that this assertion is correct for any
timelike trajectory in the expansion (40) for which the coefficient $b>0$.

The integration contour in integrals (63),(64) can be turned on the negative 
imaginary axis going around the infrared singularity at zero (in Bose-case)
along the arc with radius $\varepsilon$. Then we obtain
\begin{equation}
{\rm tr}\,\alpha^B=\frac{1}{2\pi}[-\frac {\pi}{2}-i\left(\ln\frac{4w_0}{\nu}-
\int_0^\infty dt\,\ln t\,B'(t)\right)],\qquad \nu\ll w_0,
\end{equation}                                
\begin{equation}
{\rm tr}\,\alpha^F=-\frac{1}{2\pi}i\int_0^\infty \frac{dt}{\sqrt{t}}\left(\Gamma
(\frac12+t)\,e^{t-t\ln t}-\sqrt{2\pi}\right)=\frac{1}{2\pi}i\cdot 0,8843\ldots .
\end{equation}
In the integral of the formula (68) the function $B'(t)$ is the derivative of 
the function $B(t)=\Gamma (1+t)\,e^{t-t\ln t}-\sqrt{2\pi t}$. Numerical value 
of this integral is $2,2194\ldots $. If one transforms the imaginary part of 
(68) to the form of the imaginary part of (58), then we obtain
$$ 
\ln\frac{4w_0}{\nu}-2,2194\ldots =\ln\frac{2w_0}{\gamma\nu}-0,9491\ldots .
$$
So, the ${\rm tr}\,\alpha^{B,F}$ for the exponential and hyperbolic motions
rather close to each other.

\section{Ultraviolet and infrared singularities of ${\rm tr}\,\alpha^{B,F}$}
It is not difficult to obtain the general expression for the ${\rm tr}\,\alpha^
{B,F}$ in the form of a double integral which is a functional of the mirror's
trajectory and the tangent to it at the point $u=v=0$. Indeed, after 
substituting the Bogoliubov coefficients
\begin{equation}
\alpha^B_{\omega'\omega}=\sqrt{\frac{\omega'}{\omega}}\int_{-\infty}^\infty
dv\,e^{i\omega' v-i\omega g(v)},\qquad \alpha^F_{\omega'\omega}=\int_{-\infty}
^\infty dv\,\sqrt{g'(v)}\,e^{i\omega' v-i\omega g(v)}
\end{equation}
to the formula (43) and trivial integration over the frequency $\omega'$, we
obtain
\begin{equation}
{\rm tr}\,\alpha^{B,F}=\frac{1}{2\pi}\int_0^\infty d(\frac{\omega}{\varkappa})
\int_{-\infty}^\infty dx\,\{1,\,\sqrt{G'(x)}\}\,e^{-i\frac{\omega}{\varkappa}
(G(x)-x)},
\end{equation}
where $1$ and $\sqrt{G'(x)}$ in the braces refer to the Bose- and Fermi-cases
respectively. 
The Lorentz-invariance of these expressions is evident. However, the integral 
over $(\omega/\varkappa)$ diverges on the upper limit since its integrand 
behavies as $\sqrt{\varkappa/\omega}$ at $\omega/\varkappa \to \infty$. Indeed,
for $\omega/\varkappa \to \infty$ in the integral over $x$ the $\vert x\vert
\ll 1$ will be essential. Then the functions $G(x)-x$ and $G'(x)$ can be 
replaced by the first terms of their expansions near zero, that is by $bx^2$
and $1$, see (40). Consequently, at $\omega/\varkappa \to \infty$ the integral
over $x$ is reduced to
\begin{equation}
\int_{-\infty}^\infty dx\,e^{-i\frac{\omega}{\varkappa}bx^2}=\sqrt{\frac{\pi
\varkappa}{ib\,\omega}}
\end{equation}
both in the Bose- and Fermi-case.

It is easy to show that the next term of the asymptotical expansion of the 
integral over $x$ behaves as $(\varkappa/\omega)^{3/2}$. Then, after subtraction 
from the integral over $x$ of the first term its asymptotical expansion in the 
parameter $\omega/\varkappa \to \infty$, we make the integral over 
$\omega/\varkappa$ convergent on the upper limit. If one goes from the variable
$\omega/\varkappa$ to the variable $z$,
\begin{equation}
\frac{\omega}{\varkappa}=\sqrt{\frac{\omega\omega'}{\varkappa\varkappa'}}=
\frac{b\rho}{2w_0}=\frac12 bz,
\end{equation}
the subtractive term in ${\rm tr}\,\alpha^{B,F}$ acquires the universal form
\begin{equation}
\frac{1}{2\pi}\int_0^\infty dz\,\sqrt{\frac{\pi}{2iz}}. 
\end{equation}
Remember that $z=\rho/w_0$ has the sense of the invariant momentum transfer
in units of proper acceleration.

Though the expressions
\begin{equation}
{\rm tr}\,\alpha^{B,F}=\frac{1}{2\pi}\int_0^\infty ds\lbrack \int_{-\infty}
^\infty dx\,\{1,\,\sqrt{G'(x)}\}\,e^{-is(G(x)-x)}-\sqrt{\frac{\pi}{ibs}}\rbrack ,
\quad s=\frac{\omega}{\varkappa},
\end{equation}
do not contain the ultraviolet divergences, they can contain infrared 
divergences, if the spectral function (the function of $s$ in square brackets 
of (75)) has the singlar behaviour $\propto 1/s$ for $s\to 0$. It is clear that
the behaviour of the spectral function near $s=\omega/\varkappa =0$ and also in
the main forming region of the integral over $s$ is defined by the behaviour of
the trajectory $G(x)$ far from the point of contact, where the expansion (40)
can not be applied, i.e. at the distances $\vert x\vert \gtrsim 1$.

Let us demonstrate the working of the formula (75) on an example of another one 
trajectory
\begin{equation}
u^{mir}=-\frac{1}{\varkappa}\ln\,(2-e^{\varkappa' v}),\qquad
G(x)=-\ln\,(2-e^x),
\end{equation}
for which the spectral function can be expressed in terms of the well known
transcendental functions. This trajectory, as the hyperbolic one (53), has two
asymptoties but snuggle up to them not as a power but an exponential manner. 
Therefore, on the both ends of the trajectory the proper acceleration 
\begin{equation}
a(v)=-\sqrt{\frac{\varkappa\varkappa'}{e^{\varkappa'v}(2-e^{\varkappa' v})}}
\end{equation}
tends to $-\infty $ and at zero point attains the minimal in modulus value
$a_0=-\sqrt{\varkappa\varkappa'}$.

The integral over $x$ in (75), in which the upper limit for the trajectory (76) 
is equal to $\ln\,2$, after changing the variable $x$ on $t=1-e^x$ is reduced 
to the tabular integral 2.2.5.1 of the reference book [10]. As a result we 
obtain
\begin{equation}
{\rm tr}\,\alpha^B=\frac{1}{2\pi}\int_{\varepsilon}^\infty ds\,
\lbrack\frac{\sqrt{\pi}\,\Gamma (is)}{\Gamma (\frac12+is)}-
\sqrt{\frac{\pi}{is}}\rbrack ,
\end{equation}
\begin{equation}
{\rm tr}\,\alpha^F=\frac{1}{2\pi}\int_0^\infty ds\,\lbrack\frac{\sqrt{\pi}\,\Gamma
(\frac12+is)}{\Gamma (1+is)}-\sqrt{\frac{\pi}{is}}\rbrack .  
\end{equation}

Since the spectral function has in Bose-case the infrared singularity the
corresponding divergency of the integral over $s$ for the ${\rm tr}\,\alpha^B$
is removed by introducing the small but finite lower limit 
$\varepsilon=\nu/2w_0$. Its physical meaning is the minimal momentum transfer in
units of acceleration at zero point.

After the turning of the integration contour over $s$ on the negative imaginery           
semiaxis with the detour (in Bose-case) of the singularity at zero along the 
arc of a circle with radius $\varepsilon$ we obtain
\begin{equation}
{\rm tr}\,\alpha^B=\frac{1}{2\pi}\lbrack -\frac{\pi}{2}-i(\ln\frac{2w_0}{\nu}-
B)\rbrack,
\end{equation}
\begin{equation}
{\rm tr}\,\alpha^F=\frac{1}{2\pi}i\cdot F,
\end{equation}
where the positive constants $B,\,F$ are defined by the integrals
\begin{equation}
B=\int_0^\infty dt\,\ln t\,B'(t)=1,887789\ldots\,,\qquad
B(t)=\frac{\sqrt{\pi}\Gamma (1+t)}{\Gamma (\frac12+t)}-\sqrt{\pi t},
\end{equation}
\begin{equation}
F=-\int_0^\infty dt\,\lbrack \frac{\sqrt{\pi}\Gamma (\frac12 +t)}{\Gamma (1+t)}
-\sqrt{\frac{\pi}{t}}\rbrack =1,869957\ldots\,.
\end{equation}
The imaginary part of (80) can be transformed to the form of the imaginery part 
of (57):
$$
\ln\frac{2w_0}{\nu}-1,887789\ldots =\ln\frac{2w_0}{\gamma\nu}-1,310574\ldots\,.
$$
                
The expressions for $\pm{\rm tr}\,\alpha^{B,F}$ obtained for the three different
mirror's trajectories are close to each other qualitatively and quantitatively,
see (58-59), (68-69) and (80-81). All of them have negative imaginery part which
in the Bose-case has infrared logarithmic singularity. This singularity is 
accompanied by the appearance of the real negative part for ${\rm tr}\,
\alpha^B$, namely, ${\rm Re\,tr}\,\alpha^B=-1/4$, whereas ${\rm Re\,tr}\,
\alpha^F=0$. Similar expressions for $\pm{\rm tr}\,\alpha^{B,F}$ are typical for 
the trajectories $G(x)$-function of which increases stronger (falls weaker) than
$x$ when $x$ tends to upper (lower) limit.      

Since the functionals $\pm{\rm tr}\,\alpha^{B,F}$ have, according to (52), the 
meaning of action, compare them with the changes $\Delta W_{1,0}$ of 
selfactions of electric and scalar charges at hyperbolic motion [11,12]:
\begin{equation}
\Delta W_{1,0}=-(\tau_2-\tau_1)\cdot\Delta m_{1,0},
\end{equation}
\begin{equation}
\Delta m_1=\frac{e^2w_0}{4\pi^2}[-\frac{\pi}{2}-i\left(\ln\frac{2w_0}{\gamma\mu}
-\frac12\right)]\,,\qquad \Delta m_0=-i\frac{e^2w_0}{8\pi^2}\,.
\end{equation}
In such a motion the proper acceleration of a charge is constant and the square
of the interval between two points on the trajectory is the function only of the
length of the arc connecting them:
\begin{equation}
\left(x_\alpha(\tau)-x_\alpha(\tau')\right)^2=f(\tau-\tau')\,.
\end{equation}
Therefore, the change of the charge's selfinteraction is proportional to the
duration $\tau_2-\tau_1$ of the charge's stay at hyperbolic motion multiplied 
by the mass shift $\Delta m_{1,0}$ of the charge. The mass shift owes its origin 
to the change of the interaction of a charge with its own field, which 
essentially modified at the distances $\sim w^{-1}_0$ from the charge due 
to acceleration. In other words, the shift is formed on the arclength 
$\vert \tau-\tau'\vert\sim w^{-1}_0$ with the center $\tau_c$ in any point of 
the trajectory inside the interval $(\tau_1,\,\tau_2 )$ of acceleration. The 
independence of the shift from $\tau_c$ means that it is a constant of motion. 
This is not so for the trajectories with variable acceleration, see Section 9.

As distinct from $\Delta W_{1,0}$ describing the change of interaction of the 
charge with itself due to acceleration, the functionals $\pm{\rm tr}\,
\alpha^{B,F}$ describe the interaction of accelerated mirror with the probe 
executing uniform motion along the tangent to the mirror's trajectory at the 
point where mirror has acceleration $w_0$. This interaction is transmitted by
the vector or scalar perturbations created by the mirror in the vacuum of Bose- 
or Fermi-field and carring the spacelike momentum of the order of $w_0$. 
According to (51), the field of these perturbations decreases at the distances
of the order of $w^{-1}_0$ from the mirror, exponentially in timelike 
directions and oscillating with damped amplitude in spacelike directions.
It can be said that such a field moves together with the mirror and is its 
"proper field".  Hence, the probe interacts with the mirror for a time of the 
order of $w^{-1}_0$ while the charge all the time interacts with itself and 
feels the change of interaction over the all time of acceleration. Therefore, 
it is not surprising that the expressions for $\pm{\rm tr}\,\alpha^{B,F}$ 
coincide in essence with $\Delta W_{1,0}$ if in these latter one puts $\tau_2-
\tau_1=2\pi/w_0,\;e^2=1$ and changes the sign on the opposite one. In other 
words, the $\pm{\rm tr}\,\alpha^{B,F}$ are the mass shifts of the mirror's 
proper field multiplied by characteristic proper time of their formation.    

\section{Mass shifts of electric and scalar charges at exponential motion}

For calculation of the self-actions of electric and scalar charges at 
exponential motion let us make use of the formula (30). The charge's trajectory 
(60) is convenient to use in a form of a function of proper time:
\begin{equation}
u^{mir}(\tau)=-\frac{2}{\varkappa}\ln\,(1-w_0\tau),\quad
v^{mir}(\tau)=\frac{1}{\varkappa'}(2w_0\tau-w_0^2\tau^2).
\end{equation}
Then
\begin{equation}
\dot x_{\alpha}(\tau)\dot x^{\alpha}(\tau')=-\frac{1+z^2}{1-z^2},\quad
(x-x')^2=-(\tau-\tau')^2\frac{{\rm Arth}\,z}{z},\quad
z=\frac{w_0(\tau-\tau')}{2-w_0(\tau+\tau')}.
\end{equation}

Introduce now the new variables $\xi =(\tau+\tau')/2,\;z$ instead of $\tau,\;
\tau'$. At fixed $\xi$, lying in the interval $-\infty <\xi <w_0^{-1}$, the 
variable $z$ changes in the interval $-1<z<1$. By using the causal function
$\Delta^f_4$ expressed via Macdonald's function, we obtain
\begin{eqnarray}
\Delta W_1=e^2\int_{-\infty}^{w_0^{-1}}d\xi\,(\frac{1}{w_0}-\xi)\,\int_{-1}^1
dz\,\dot x_{\alpha}(\tau)\dot x^{\alpha}(\tau')\,\Delta^f_4(x-x',\mu)\vert^F_0=
\nonumber\\
-\frac{e^2}{2\pi^2}\int_{-\infty}^{w^{-1}_0}d\xi\,\int_0^\infty\frac{du\,\mu}
{{\rm sh}\,2u}\lbrace{\rm ch}\,2u\frac{{\rm th}u}{u}\,K_1(i\lambda\sqrt{u\,
{\rm th}\,u})-K_1(i\lambda\,{\rm th}\,u)\rbrace.                     
\end{eqnarray}
In the last expression instead of $z$ the variable $u={\rm Arth}\,z$ is used,
and $\lambda$ is the function of $\xi$, $\lambda (\xi)=2\mu (w_0^{-1}-\xi)$.

Our problem now is to find the integral over $u$ in that region of the variable 
$\xi$ where $\lambda (\xi)\ll 1$, supposing, of course, that the infrared
parameter $\mu/w_0\ll 1$. This integral coinsides, in essense, with the mass 
shift of an electric charge
\begin{equation}
\Delta m_1=\frac{e^2}{2\pi^2}\int_0^\infty\frac{du\,\mu}{{\rm sh}\,2u}
\lbrace{\rm ch}\,2u\,\sqrt{\frac{{\rm th}\,u}{u}}K_1(i\lambda
\sqrt{u\,{\rm th}u})-K_1(i\lambda\,{\rm th}u)\rbrace.
\end{equation}
To calculate $\Delta m_1$ when $\lambda (\xi)\ll 1$ let us divide the 
integration interval into two intervals, $0\leqslant u\leqslant u_1$ and 
$u_1\leqslant u<\infty$, by the point $u_1$ where $u_1\gg 1$, but 
$\lambda u_1\ll 1$. Then, using the expansion of Macdomald's function at small 
argument, we obtain
\begin{eqnarray}     
\Delta m_1\approx \frac{e^2w_0}{4\pi^2(1-w_0\xi)}\lbrace \frac1i\int_0^{u_1} du
\,\left(\frac{{\rm cth}\,2u}{u}-\frac{1}{2\,{\rm sh}^2 u}\right)+\int_{u_1}^\infty
\frac{du\,\lambda}{\sqrt{u}}\,K_1(i\lambda u)\rbrace=  \nonumber\\
\frac{e^2w_0}{4\pi^2(1-w_0\xi)}\lbrace -\pi-i\left(2\ln\frac{w_0}{\gamma\mu
(1-w_0\xi)}+\ln\frac{2\gamma}{\pi}+\frac12\right)\rbrace.
\end{eqnarray}

The mass shift $\Delta m_0$ of the scalar charge differs from (90) by the 
replacement ${\rm ch}\,2u\to -1$ in the first term of the braces and by the 
change of the sign of the second term. Then at the same condition
$\lambda (\xi)\ll 1$ we get      
\begin{equation}
\Delta m_0=-i\frac{e^2w_0}{4\pi^2(1-w_0\xi)}(\ln\,2-\frac12).
\end{equation}

As is seen from (91), (92) and (66), the mass shift depends on the absolute
value $w(\xi)=w_0/(1-w_0\xi)$ of the proper acceleration of the mirror at the 
instant $\xi$, which may be considered as a center of forming region of the 
shift. As the acceleration essentially changes on such an interval, the mass
shifts (91), (92) do not coincide with the mass shifts (84), (85) of uniformly-
accelerated charges if one replaces $w(\xi)$ by $w_0$. Nevertheless, rather 
close coincidence arises at the replaces $w(\xi)\to 0,5\,w_0$ and $w(\xi)\to
2,6\,w_0$ for the $\Delta m_1$ and $\Delta m_0$ correspondingly.

\section{Conclusion}

The basis for the symmetry between the processes induced by the mirror in
two-dimentional and by the charge in four-dimentional space-time is the relation
(14), (15) between the Bogoliubov's coefficients $\beta^{B,F}_{\omega'\omega}$ 
and the current density $j^\alpha (k)$ or charge density $\rho (k)$ depending 
on the timelike momentum $k^{\alpha}$. The squares of these quantities represent 
the spectra of real pairs and particles radiated by accelerated mirror and 
charge.

In the present paper the symmetry is extended to the selfactions of the mirror 
and the charge and to the corresponding vacuum-vacuum amplitudes, cf. (29) and 
(30). In essence, it is embodied in the discovered relation (20) between 
propagators of a massive pair in two-dimentional space and of a single particle 
in four-dimentional space.

The formula (29) for $W^{B,F}$ was obtained provided that the mean number 
${\rm tr}\,\beta^{+}\beta$ of pairs created is small and the interference of
two or more pairs is negligible. In the general case the $W^{B,F}$ is given by
the formula (27), which can be written also in the form
\begin{equation}
2\,{\rm Im}\,W^{B,F}=\pm{\rm tr}\,\ln (\alpha^{+}\alpha)^{B,F},
\end{equation}
since $\alpha^+\alpha \mp\beta^+\beta=1$, see [7], [4]. As is seen from (27) or
(93) the imaginery part of the action differs from zero and then is positive 
only if $\beta \ne 0$, i.e. if the radiation of real particles is happened 
indeed.

Formula (93) allows to choose for $W^{B,F}$ the expression
\begin{equation}
W^{B,F}=\pm i\,{\rm tr}\,\ln\,\alpha^{B,F},
\end{equation}
that was named natural by DeWitt [7]. However, this expression is by no means 
unique. The expressions $W^{B,F}=\pm i\,{\rm tr}\,
\ln (\alpha e^{i\gamma})^{B,F}$ and 
$W^{B,F}=\pm i\,{\rm tr}\,\ln\,\alpha^{B,F +}$
have the same imaginery part. Nevertheless, the formula (94) is interesting as
the definition both the real and imaginery parts of the selfactions $W^{B,F}$ by
means of the Bogoliubov's coefficients $\alpha^{B,F}_{\omega'\omega}$ only,
which, according to the formulae (14), (15), reduce to the current density
$j^\alpha (q)$ or to the charge density $\rho (q)$ dependent on the spacelike 
momentum $q^\alpha$. This means that the field of the corresponding perturbations 
propagates in vacuum together with the mirror, comoves it, and, at the same 
time, it containes the information about the radiation of the real quanta.

Unfortunately, the author failed to find a simple integral representation for
the matrix $\ln\,\alpha$. Nevertheless, if one again assumes that the mean 
number of emitted particles is small, then one may consider $\alpha$ or
$i\alpha$ close to 1. Then, expanding the $\ln\,i\alpha$ near $i\alpha=1$ and
confine ourselves by the first term we obtain
\begin{equation}
W^{B,F}=\pm i\,{\rm tr}\,\ln\,i\alpha^{B,F}\approx
\pm i{\rm tr}\,(i\alpha^{B,F}-1)=\mp{\rm tr}\,\alpha^{B,F}+\ldots\;.    
\end{equation}
These qualitative arguments allow to state that the functionals $\pm{\rm tr}\,
\alpha^{B,F}$ are similar to the corresponding selfactions with opposite sign
and therefore must have the negative imaginery parts. This is confirmed by all
examples considered in Sections 7 and 8. Nevertheless, the exact physical 
meaning of the quantities $\pm{\rm tr}\,\alpha^{B,F}$ is clearly defined by the
formula (52).

Here we want also to concentrate attention on one prediction followed from the 
symmetry between processes induced by the mirror in two-dimentional and by 
the charge in four-dimentional space-times. The symmetry predicts the value
$e_0^2=1$ for the charge squared (in Heviside's units), that corresponds to the 
fine structure constant $\alpha_0=1/4\pi$. Since the radiation corrections are
not taken into account in both spaces, and the processes in 1+1-space are due 
to the purely geometrical boundary condition, it is natural to think that the
above mentioned values for the charge squared and for the fine structure 
constant are the nonrenormalized bare values of these constants. Therefore they 
are marked by index 0.

It is very interesting that the bare fine structure constant has the purely 
geometrical origin and, also, that its value is small: $\alpha_0=1/4\pi\ll 1$.
The smallness of $\alpha_0$ has the essential meaning for the quantum 
elecrodynamics where it justifies a priori the applicability of the perturbation 
theory and where the radiative corrections in accordance with the well known 
formula [13, 14]
\begin{equation}
\alpha=\frac{\alpha_0}{1+\frac{\alpha_0}{3\pi}\,N\,\ln\frac{\Lambda^2}{m^2}}
\end{equation}
decrease the renormalized value of $\alpha$ in comparison with unrenormalized 
one. Here $N$- is the number of charged particles with masses in the interval
$(m,\Lambda)$, and $\Lambda$ is the upper limit of the particle energy up to 
which the quantum electrodynamics is correct.

\section{Appendix}
The singular function $\Delta^{LR}_d(z,\nu)$ in d-dimentional space-time, as 
well as the causal function $\Delta^f_d(z,\mu)$, is convenient to define by 
Fourier representation 
\begin{equation}
\Delta^{LR}_d(z,\nu)=\int\frac{d^dq}{(2\pi)^d}\,\frac{e^{iqz}}{q^2-\nu^2+
i\varepsilon},\qquad \Delta^f_d(z,\mu)=\int\frac{d^dq}{(2\pi)^d}\,\frac{e^{iqz}}
{q^2+\mu^2-i\varepsilon}.
\end{equation}
These functions are the even singular solutions of the inhomogeneous wave 
equations
\begin{equation}
(-\partial^2-\nu^2)\,\Delta^{LR}(z,\nu)=\delta (z),\qquad 
(-\partial^2+\mu^2)\,\Delta^{f}(z,\mu)=\delta (z),
\end{equation}
with opposite signs before the parameters $\nu^2$ and $\mu^2$, where $\nu$ and
$\mu$ are the momentum transfer and mass.  Their proper time representations
(in particular, for $d$=4)
\begin{equation}
\Delta^{LR}_4(z,\nu)=\frac{1}{(4\pi)^2}\int_0^\infty \frac{ds}{s^2}\,e^{-i\nu^2
s-iz^2/4s},\qquad \Delta^{f}_4(z,\mu)=\frac{1}{(4\pi)^2}\int_0^\infty \frac{ds}
{s^2}\,e^{-i\mu^2s+iz^2/4s},
\end{equation}
as well as the explicit expressions in terms of Macdonald's function, differ by 
the complex conjugation and the replacement $\mu\to i\nu $ or by the 
replacement $z^2\to-z^2,\;\mu\to\nu$.

For the symmetry being discussed in this paper the integral relation 
\begin{equation}
\Delta^{LR}_{d+2}(z,\nu)=-\frac{1}{4\pi}\int_{\nu^2}^\infty d\rho^2\,\Delta^{LR}
_d(z,\rho)
\end{equation}
is very important. It differs from the similar relation (20) for the causal 
functions not only by the sign. Being written for $z^2<0$, it is understood
for $z^2>0$ in a sense of analytical continuation in the lower half-plane of
complex $z^2$. Whereas the relation (20), being written for $z^2>0$, is 
understood for $z^2<0$ as the analytical continuation in the upper half-plane 
of complex $z^2$. For the $\Delta^{+}$-functions such a continuation must be
carry out in the upper half-plane if $z^0>0$, and in the lower one if $z^0<0$.

The author is grateful to A.I. Nikishov and A.I. Ritus for valuable discussions
and help. The work was carried out with financial support of Russian
Fund for Fundamental Research (Grants 00-15-96566 and 02-02-16944).

\end{document}